\documentclass[twocolumn,aps,prl,superscriptaddress,showpacs,floatfix]{revtex4}
\usepackage[latin9]{inputenc}
\usepackage{mathtools}
\usepackage{amsmath}
\usepackage{graphicx}
\usepackage{esint}

\makeatletter
\@ifundefined{textcolor}{}
{%
 \definecolor{BLACK}{gray}{0}
 \definecolor{WHITE}{gray}{1}
 \definecolor{RED}{rgb}{1,0,0}
 \definecolor{GREEN}{rgb}{0,1,0}
 \definecolor{BLUE}{rgb}{0,0,1}
 \definecolor{CYAN}{cmyk}{1,0,0,0}
 \definecolor{MAGENTA}{cmyk}{0,1,0,0}
 \definecolor{YELLOW}{cmyk}{0,0,1,0}
}

\usepackage{bm}

\makeatother

\begin{document}
%#############################################################################################

\title{An efficient thermal diode with ballistic spacer}

\author{Shunda Chen}

\email{shdchen@ucdavis.edu}

\affiliation{Department of Chemistry, University of California Davis, One Shields
Ave. Davis, CA 95616, USA}

\author{Davide Donadio}

\email{ddonadio@ucdavis.edu}

\affiliation{Department of Chemistry, University of California Davis, One Shields
Ave. Davis, CA 95616, USA}

\affiliation{Ikerbasque, Basque Foundation for Science, E-48011 Bilbao, Spain}

\author{Giuliano Benenti}

\email{giuliano.benenti@uninsubria.it}

\affiliation{Center for Nonlinear and Complex Systems, Dipartimento di Scienza e
Alta Tecnologia, Universit\`a degli Studi dell'Insubria, via Valleggio 11, 22100 Como, Italy}

\affiliation{Istituto Nazionale di Fisica Nucleare, Sezione di Milano, via Celoria
16, 20133 Milano, Italy}

\affiliation{NEST, Istituto Nanoscienze-CNR, I-56126 Pisa, Italy}

\author{Giulio Casati}

\email{giulio.casati@uninsubria.it}

\affiliation{Center for Nonlinear and Complex Systems, Dipartimento di Scienza e
Alta Tecnologia, Universit\`a degli Studi dell'Insubria, via Valleggio 11, 22100 Como, Italy}

\affiliation{International Institute of Physics, Federal University of Rio Grande do Norte,
Campus Universit\'ario - Lagoa Nova, CP. 1613, Natal, Rio Grande Do Norte 59078-970, Brazil}

\date{\today}
\begin{abstract}
Thermal rectification is of importance not only for fundamental physics,
but also for potential applications in thermal manipulations and thermal management.
However, thermal rectification effect usually decays rapidly with system size.
Here, we show that a mass-graded system, with two diffusive leads separated 
by a ballistic spacer, can exhibit large thermal rectification effect, with
the rectification factor independent of system size.
The underlying mechanism is explained in terms of the effective size-independent thermal gradient and the match/mismatch of the phonon bands.
We also show the robustness of the thermal diode upon variation of the 
model's parameters. Our finding suggests a promising way for designing realistic efficient thermal diodes.

\end{abstract}

\pacs{05.70.Ln; 05.60.-k; 44.10.+i}

%05.70.Ln 	Nonequilibrium and irreversible thermodynamics 
%05.60.-k 	Transport processes
%44.10.+i 	Heat conduction

\maketitle

%\section{Introduction}
\textit{Introduction.}
Heat is a ubiquitous form of energy, of which we have limited control. 
Waste heat limits the performance of the smallest chips present in 
electronic devices, such as laptop computers and cellular phones. 
Energy supply and cooling is also a challenge for the 
large data centers and supercomputers. Therefore, efficiently harnessing 
thermal energy would have an enormous societal impact.
Effective control of heat currents requires the development of a 
new class of nanoscale thermal devices, namely thermal rectifier and 
amplifiers, analogous to electronic diodes and transistors. 

Nonlinear dynamics indicates possible pathways toward thermal 
diodes and transistors \cite{Casati1,Casati2,Casati3,BaowenRMP,Benenti2016}.
However, in spite of pioneering experimental 
investigations of thermal rectification of phonon transport
\cite{majumdar,terasaki1,terasaki2,terasaki3,NJP11,tian12,wang17},
a satisfactory real-life implementation of these concepts 
has not been achieved yet \cite{giazotto}.

The rectification factor can be defined as 
\begin{equation}
f_{r}=\frac{(J_{+}-J_{-})}{J_{-}}\times100\%,
\label{eq:frectification}
\end{equation}
where $J_{+}$ and $J_{-}$ represent, respectively, the larger heat flow
and the smaller heat flow, obtained by inverting the temperature bias
applied to the system. 
While numerical simulations of theoretical models of nonlinear systems 
predict rectification factors of the order of $10000\%$, experimental 
phononic devices so far are limited to $f_r\approx 70\%$ 
\cite{majumdar,terasaki1,terasaki2,terasaki3,NJP11,tian12,wang17,giazotto}.

A general and yet unsolved problem of thermal rectifiers 
is that rectification 
rapidly decays to zero as the size increases. This effect is at first sight
unavoidable since, for a given temperature bias, the temperature gradient
decreases as the system size increases. Consequently, the linear response
regime where rectification vanishes should be approached. 
From a practical viewpoint, it would be highly desirable to overcome
this problem, since it is difficult to apply large temperature biases
on small sizes. A first proposal
\cite{pereira2013,shunda2015}
was based on mass-graded system with long-range interactions, where new channels (interactions) among different sites with different masses were created, due to the introduction of long-range interactions. The long range interactions (new channels) in mass-graded system connect distant particles with very different masses, leading to the increase of asymmetry, which favor the asymmetric flow, i.e. rectification, and avoid the usual decay of rectification with system size. The idea of introducing new channels is inspiring, and further motivates us to look for experimentally feasible ways
to have large thermal rectification effect with
the rectification factor independent of the system size. 

In this Letter, we consider a one-dimensional, segmented mass-graded system, 
with the particles in the outer parts of the system (hereafter referred
to as left and right lead) exposed to nonlinear on-site potentials,
and separated by a ballistic channel, where the interaction between the
particles is harmonic.  
We show that this model exhibits large rectification. Even
more importantly, thanks to the ballistic spacer, the size dependence
of the rectification factor is removed. That is, $f_r$
does not decay with the system size.
The rectification is explained in terms of the match/mismatch 
(depending on the direction of the applied thermal bias) of the
phonon bands for the two leads and for the ballistic channel.
We also show that our results are quite robust upon variation of 
the system parameters and that our model is flexible. 
Indeed the main ingredient for rectification is the different 
temperature dependence of the phonon bands in the two leads,
and this effects can be obtained in several ways. 
For instance, with different on-site potentials in the left and right 
leads rather than with a mass-graded system.  
Finally, we show that large rectification for large system sizes
can still be obtained if the spacer is sub-ballistic rather
than ideally ballistic. The stability of the effect upon nonlinear 
interactions in the spacer is relevant in view of possible experimental
implementations. Prospects in this regard are also discussed.

\begin{figure}[!]
\includegraphics[scale=0.166]{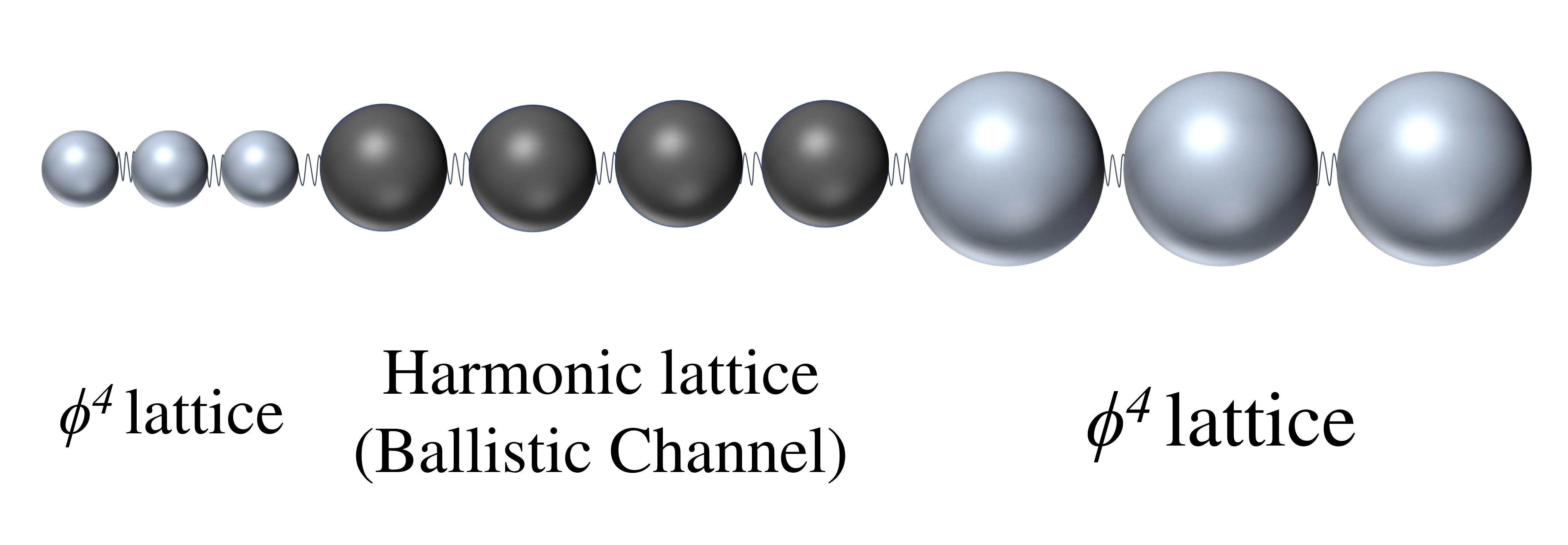} \vskip-.4cm \caption{Schematic drawing of our model. It consists of three
parts, one central ballistic channel with $N_{C}$ particles of mass
$m_{C}$, and two leads ($\phi^{4}$ lattices), with $N_{L}$ ($N_{R}$)
particles of mass $m_{L}$ ($m_{R}$) and on-site potential strength
$\gamma_{L}$ ($\gamma_{R}$) in the left (right) lead. }
\label{fig:model}
\end{figure}

\textit{The model.}
The system we will discuss here 
(schematically drawn in Fig.~\ref{fig:model}) 
is a one-dimensional chain of $N_{\rm tot}$ 
oscillators described by the Hamiltonian
\begin{equation}
H=\sum_{i=1}^{N_{\rm tot}} \left( \frac{p_i^2}{2 m_i}+
\frac{\gamma_i q_i^4}{4}\right) +
\frac{1}{2}\sum_{i=1}^{N_{\rm tot}-1}\left(q_{i+1}-q_i\right)^2,
\label{eq:Hamiltonian}
\end{equation}
where $q_i$ is the displacement from the
equilibrium position of the $i$-th particle with mass $m_{i}$ and
momentum $p_{i}$, and $\gamma_i$ measures the strength of the on-site 
quartic potential. 
The systems consists of  $N_{L}$ ($N_{R}$) particles with
mass $m_{L}$ ($m_{R}$) and strength of the on-site potential 
$\gamma_L$ ($\gamma_R$) 
in the left (right) lead. These two anharmonic leads ($\phi^4$ lattices)
are connected by a ballistic channel, that is, by a purely harmonic
central chain of $N_C$ particles with mass $m_C$ and zero 
on-site potential, $\gamma_C=0$. The total system size is $N_{\rm tot}=N_{L}+N_{C}+N_{R}$.

%%%%%%%%%%%%%%%%%%%%%%%%%%%%%%%%%%%%%%%%%%%%%%%%%%%%% figure 1

%%%%%%%%%%%%%%%%%%%%%%%%%%%%%%%%%%%%%%%%%%%%%%%%%%%

In our non-equilibrium simulations, 
two Langevin heat baths (with dissipation coefficient of 1) \cite{Dharrev}
at different temperatures $T_{-}$ and $T_{+}$ are attached to the
two ends of the leads [leftmost (rightmost) particle of left (right) lead]; we take $T_{-}=T(1-\frac{\triangle T}{T})$
and $T_{+}=T(1+\frac{\triangle T}{T})$. 
We investigate the thermal transport properties of our model,
focusing in particular on the rectification factor $f_r$,
defined in Eq.~(\ref{eq:frectification}).

In our simulations, each system is evolved for a long enough time ($>$$10^{8}$) to
ensure that it has reached the stationary state. After that the heat
current is measured. The simulation time for averaging heat flux is larger than $10^{10}$.
More specifically, we verified the average heat flow through each particle is the same (within 0.01\%).
Free boundary conditions and velocity-Verlet algorithm (with time step 0.01) are used. We
have checked simulations using fixed boundary conditions and simulations using Runge-Kutta algorithm of seventh to eighth
order, and obtained consistent results. As commonly adopted in the literature, we use dimensionless
units.

\begin{figure}[!]
\includegraphics[scale=0.32]{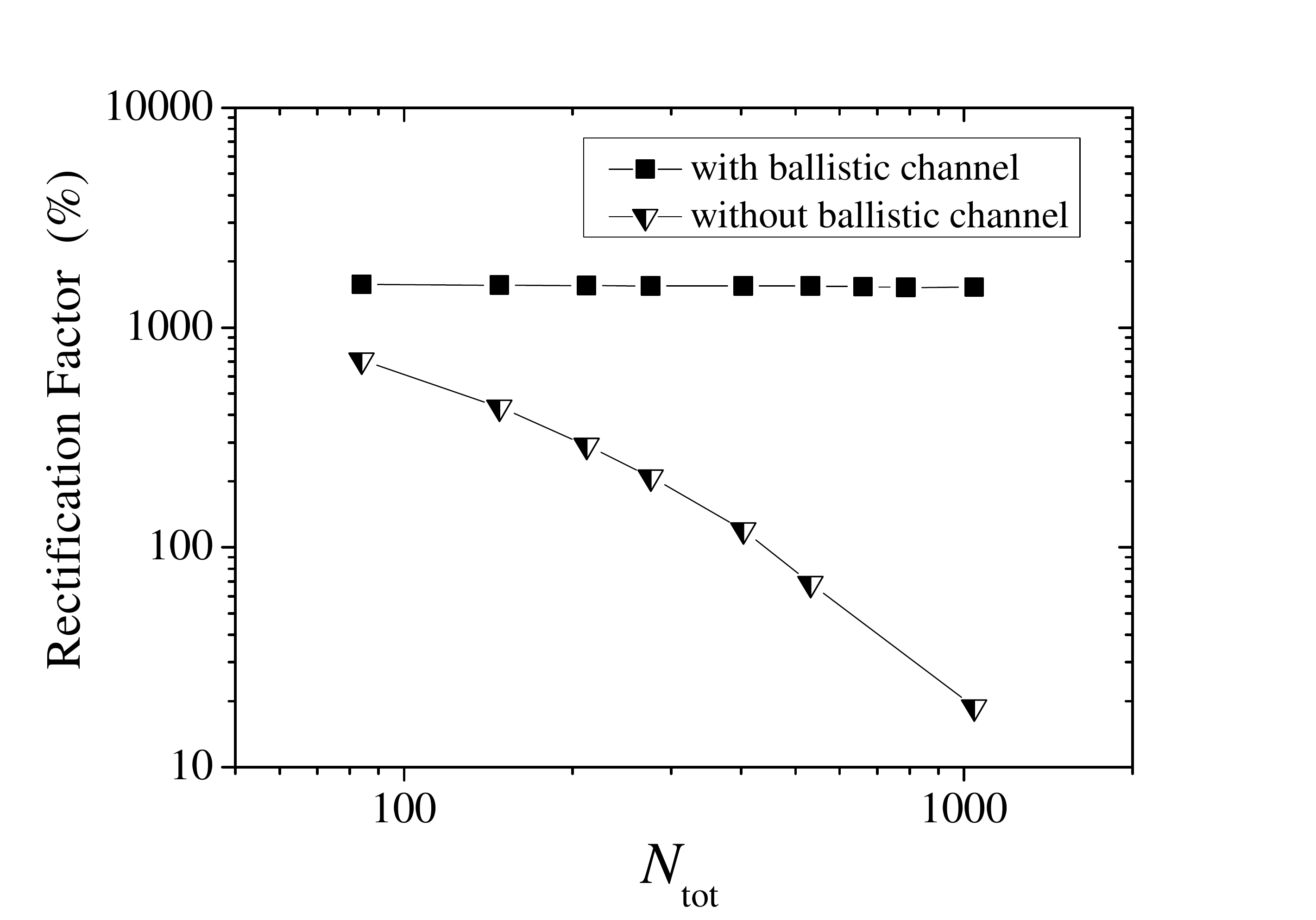} \vskip-0.2cm \caption{Thermal rectification factor $f_r$ versus system size $N_{\rm tot}$ for chains
with or without ballistic channel 
(squares are for $\phi^{4}$-harmonic-$\phi^{4}$
model, i.e. with ballistic channel, and triangles are for 
$\phi^{4}$-$\phi^{4}$-$\phi^{4}$
model, i.e. without ballistic channel). Here, $T_{+}=9.5$, $T_{-}=0.5$,
$N_{L}=N_{R}=10$, $m_{L}=1$, $m_{C}=4.5$, $m_{R}=10$, $\gamma_{L}=\gamma_{R}=1$,
and $\gamma_{C}=0$ ($\gamma_{C}=1$) 
for the model with (without) ballistic channel.}
\label{Fig:diffvsballistic}
\end{figure}

%\textit{Results.}
\textit{Size-independent rectification.}
We start our numerical studies by highlighting the role played by the ballistic
channel. For that purpose, we compare in Fig.~\ref{Fig:diffvsballistic}
the size-dependence of the rectification factor $f_r$ for 
model (\ref{Fig:diffvsballistic}), when the intermediate
channel is either diffusive ($\gamma_C=\gamma_L=\gamma_R$) or
ballistic ($\gamma_C=0$). While in the former case 
($\phi^4-\phi^4-\phi^4$ model) the rectification 
factor rapidly decays with the system size, in the 
latter case ($\phi^4$-harmonic-$\phi^4$ model)  
the rectification factor is size-independent. 
Note that we found the larger heat flow $J_+$ when the higher temperature
$T_+$ was on the side of the heavier masses as in previous work \cite{majumdar,shunda2015}. 

%%%%%%%%%%%%%%%%%%%%%%%%%%%%%%%%%%%%%%%%%%%%%%%%%%%%%%% figure 2

%%%%%%%%%%%%%%%%%%%%%%%%%%%%%%%%%%%%%%%%%%%%%%%%%%

\begin{figure}[!]
\includegraphics[scale=0.30]{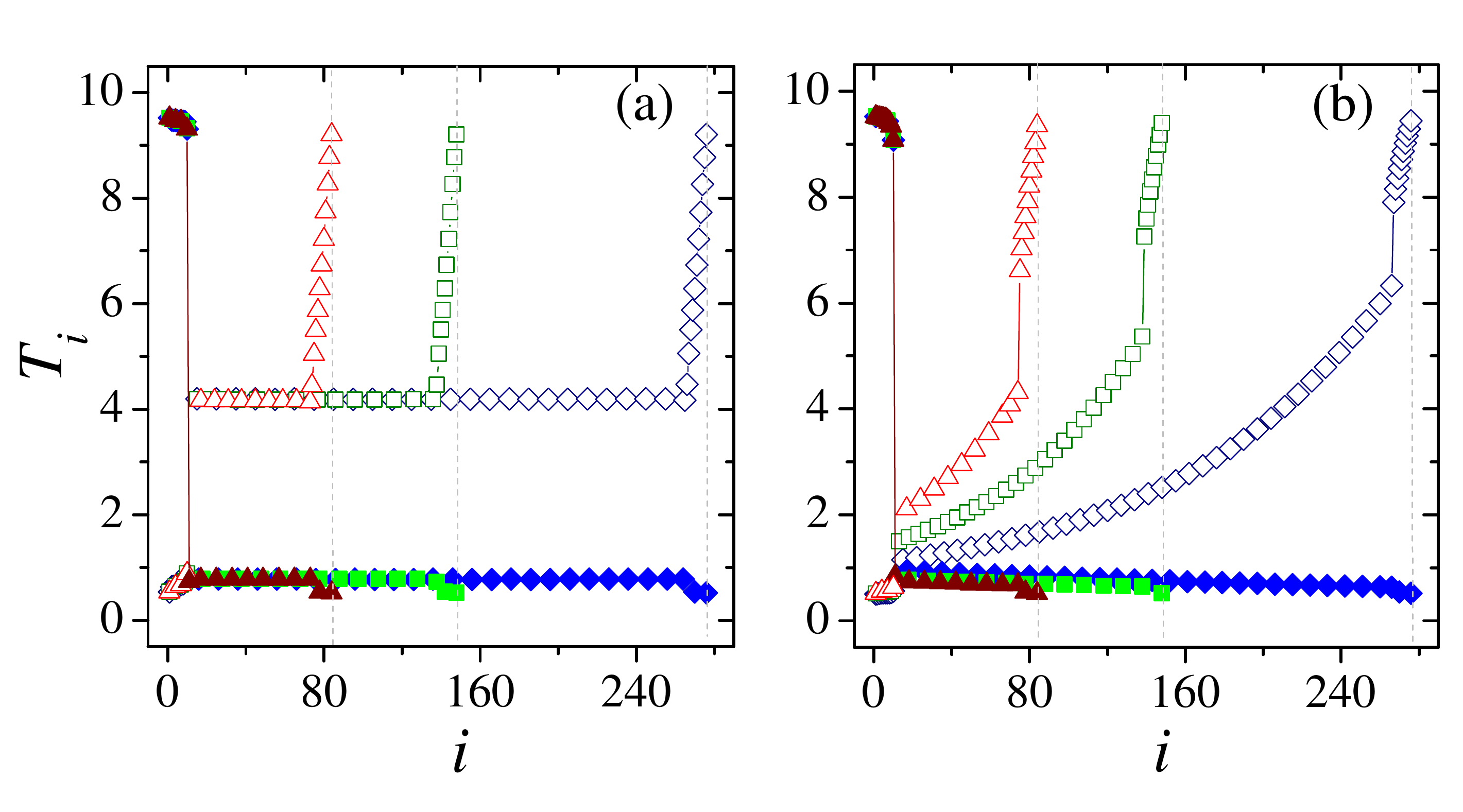} \vskip-.4cm
\caption{(Color online) (a) Temperature profiles, both for forward 
(left-right, full symbols) and 
backward (right-left, empty symbols) temperature bias, for 
the model with a ballistic channel ($\phi^{4}$-harmonic-$\phi^{4}$ model)
and (b) for the model without a ballistic channel 
($\phi^{4}$-$\phi^{4}$-$\phi^{4}$ model).
In both cases $N_{\rm tot}=84$ (triangles), $148$ (squares), and $276$ (diamonds),
respectively. Here $T_+=9.5$, $T_-=0.5$, 
$m_{L}=1$, $m_{R}=10$, $m_{C}=4.5$,
$N_{L}=N_{R}=10$, $\gamma_{L}=\gamma_{R}=1$, and $\gamma_C=0$
(panel (a)) or $\gamma_C=1$ (panel (b)).}
\label{Fig:profiles}
\end{figure}

In Fig.~\ref{Fig:profiles}, 
we compute the temperature profile inside the system, that is, 
the local temperature at site $i$ ($i=1,...,N_{\rm tot}$) defined as
$T_i=m_i\langle \dot{q}_i^2 \rangle$, where 
$\langle\, . \, \rangle$ stands for temporal average under
steady-state conditions. 
The different behavior of the rectification factor as a function of 
the system size with and without an intermediate ballistic channel
is accompanied by a striking difference in the local temperature 
profile. In both cases, the temperature profile exhibits a marked 
asymmetry between forward (left-right) and backward (right-left) 
temperature bias, as 
needed to have large rectification factors. 
For the model with a ballistic channel, the temperature 
profile is flat inside the central, harmonic region, with a value
of the temperature which is size-independent. This implies that also the 
temperature gradients in the two leads are size independent.
On the other hand, for the model without ballistic channel 
the temperature profile is such that the temperature
gradients reduce with the systems size. As a consequence, the 
linear response regime is approached and the rectification factor 
decreases with the system size.

%%%%%%%%%%%%%%%%%%% figure3 temperature profiles for systems with or without ballistic channel.

%%%%%%%%%%%%%%%%%%%%%%%%%%%%%%%

\begin{figure}[!]
\includegraphics[scale=0.3]{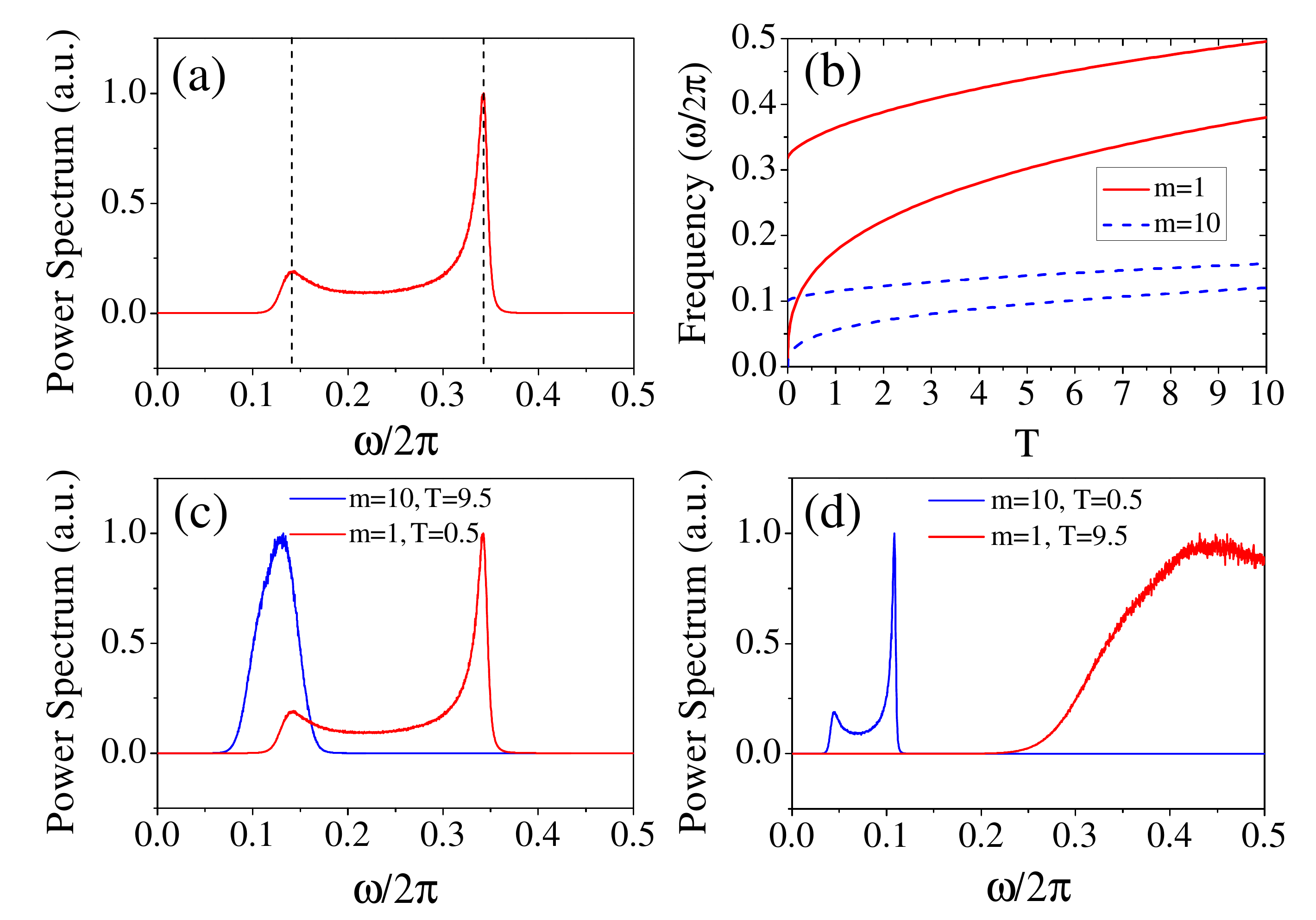} \vskip-.4cm \caption{(Color online) 
Vibrational power spectrum for the $\phi^4$ lattice,
calculated at thermal equilibrium at temperature $T$.
(a) $m=1$, $T=0.5$; the dashed lines correspond to the 
boundaries of the analytically
estimated phonon band. 
(b) Analytically estimated phonon band as a function of temperature,
for $m=1$ (red solid lines) and $m=10$ (blue dashed lines).
(c) Power spectrum for $m=10$, $T=9.5$ (blue line) and for
$m=1$, $T=0.5$ (red line). 
(d) Power spectrum for $m=10$, $T=0.5$ (blue line) and for 
$m=1$, $T=9.5$ (red line).
In all cases the nonlinearity strength $\gamma=1$.}
\label{fig:pspectrum}
\end{figure}

\textit{Rectification mechanism.}
The numerically observed thermal rectification can be explained
in terms of the match/mismatch of the phonon bands for the two 
leads, when the higher temperature is on the side of the heavier/lighter
masses.
We compute the vibrational power spectrum as the Fourier transform of the 
velocity-velocity autocorrelation function of a particle:
\begin{equation}
P(\omega)=\int\langle v(t)v(0)\rangle e^{-i\omega t}dt,
\end{equation}
where the simulations are for a closed $\phi^4$ system, with 
particle mass $m$, periodic boundary
conditions, and at thermal equilibrium.
An example of power spectrum is shown in 
Fig.~\ref{fig:pspectrum}(a).
The numerical results are in good agreement with those
obtained from the effective phonon approach \cite{destri,baowen}. 
For a $\phi^4$ lattice, this approach predicts an effective 
(i.e., renormalized by the nonlinearity) phonon spectrum 
in the band $\sqrt{1.23\,T^{2/3}/m}\le \omega \le 
\sqrt{(4+1.23\,T^{2/3})/m}$ \cite{baowen}.  
This estimated phonon band is shown as a function of temperature
in Fig.~\ref{fig:pspectrum}(b).
The temperature dependence
is stronger for light masses, while for heavy masses the band
is almost flat, since the nonlinearity in this case is weak. 
Therefore, as shown in Fig.~\ref{fig:pspectrum}(c)-(d) we have 
larger overlapping of the power spectrum for the two leads
(and consequently higher thermal conductivity)
when the higher temperature is applied to the heavier lead.
We note that a similar match/mismatch of the vibrational 
power spectrum is obtained from the nonequilibrium simulations of the entire
$\phi^4$-harmonic-$\phi^4$ model exposed to thermal baths, 
comparing the power spectrum for two particles, one in the lighter
lead and the other in the heavier lead (data not shown here).

\begin{figure}[!]
\includegraphics[scale=0.3]{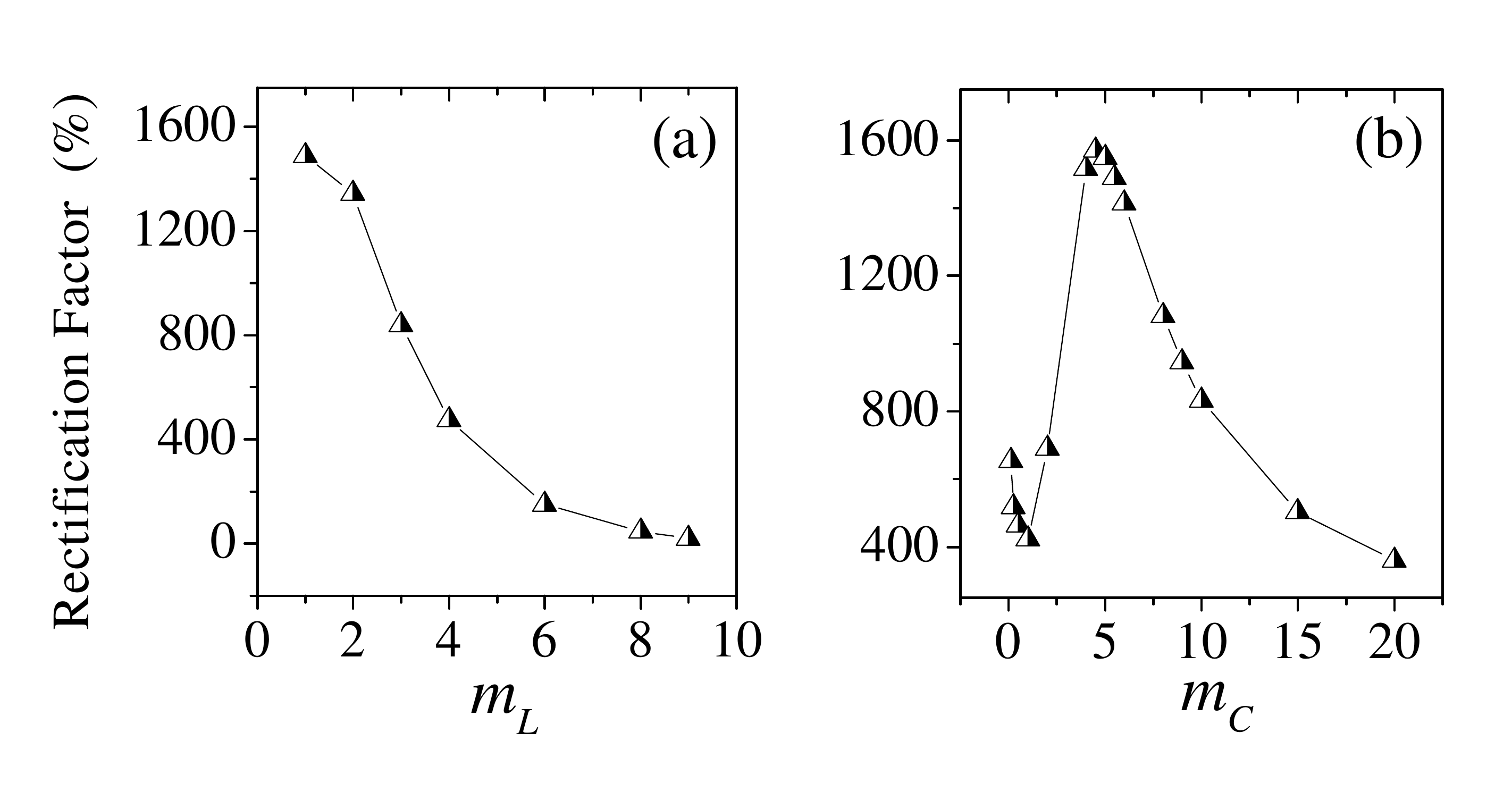} \vskip-.4cm \caption{Dependence of 
the rectification factor on: 
(a) mass gradient ($m_{R}=10$, $m_{C}=(m_{L}+m_{R})/2$, $T_{+}=9.5$,
$T_{-}=0.5$) and (b) mass of the particles in the 
ballistic channel ($m_{L}=1$,
$m_{R}=10$, $T_{+}=9.5$, $T_{-}=0.5$). 
In both panels, $N_{L}=N_{R}=10$, $N_{C}=64$,
total system size $N_{\rm tot}=84$, $\gamma_{L}=\gamma_{R}=1$.}
\label{fig:deltaTmasses}
\end{figure}

We also point out that the rectification factor is optimized when the mass
of the ballistic channel is chosen in such a way that its phonon band
has a significant overlap with the phonon bands of both leads
(for the thermal bias direction corresponding to the larger thermal 
current $J_+$). 
This happens for a value of $m_C$ intermediate between $m_L$ and 
$m_R$. More precisely, for the masses and temperatures chosen in our
simulations, we found that the optimal value for rectification
is $m_C\approx 4.5$ (see Fig.~\ref{fig:deltaTmasses}).

%%%%%%%%%%%%%%%%%%%%%%%%%%%%%%%%%%%%%%%%%%%%% figure4

%%%%%%%%%%%%%%%%%%%%%%%

\textit{Robustness of rectification.}
It is important to study the robustness of the rectification effect
upon variations of the system's 
parameters. 
In Fig.~\ref{fig:deltaTmasses}(a) we change the mass $m_L$ of the lighter
particles. We can see that, while the rectification factor is 
huge for large mass difference, $m_L/m_R\ll 1$, a significant 
effect still remains for moderate mass asymmetry.
In  Fig.~\ref{fig:deltaTmasses}(b) we show that 
a large rectification is 
observed in a broad range of values for $m_C$ around the 
optimal value $m_C\approx 4.5$, 
including the cases when $m_C$ approaches either $m_L$ or $m_R$. 

%%%%%%%%%%%%%%%%%%% figure5

%%%%%%%%%%%%%%%%%%%%%%%%%%%%%%%%%%%%%%%%%%%%%%% 

\begin{figure}[!]
\includegraphics[scale=0.30]{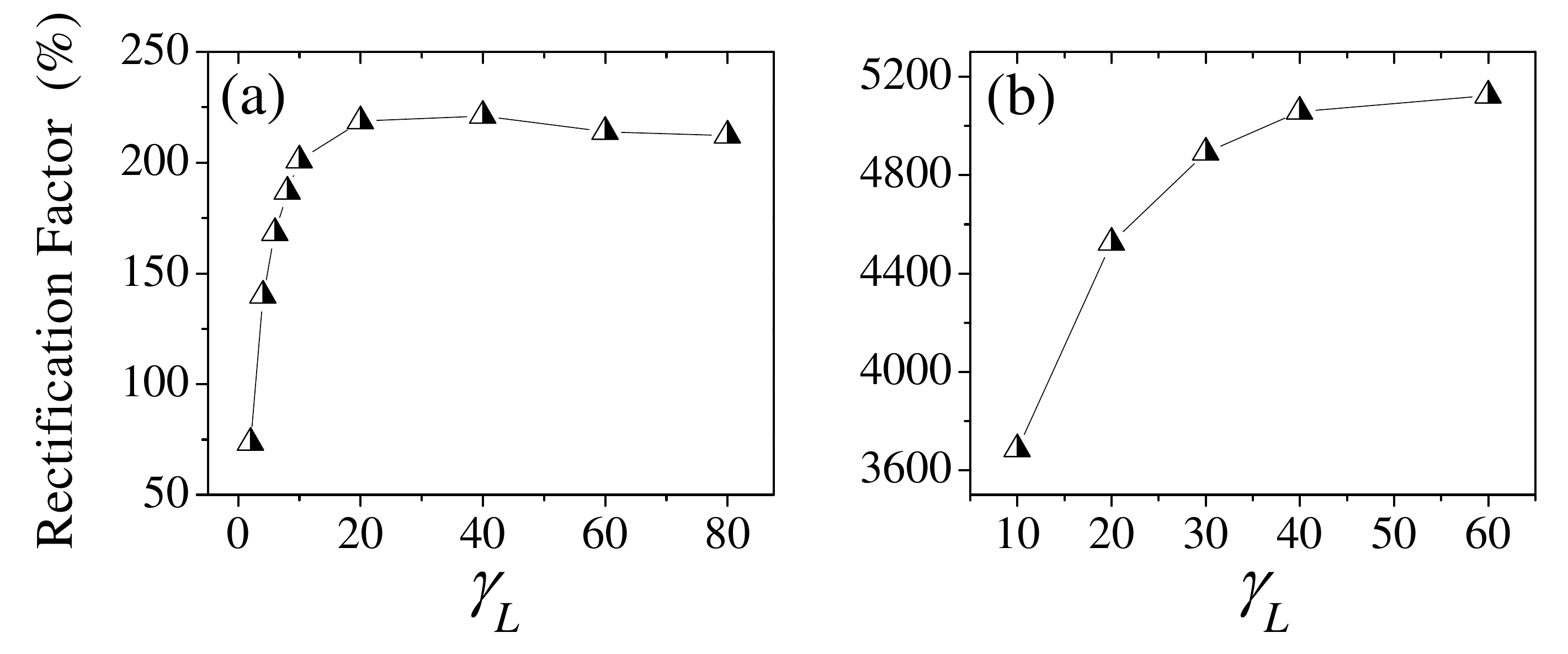} \vskip-.4cm \caption{(a) Dependence 
of the rectification factor on the strength $\gamma_L$ of the 
on-site potential in the left lead for (a) the equal-mass system
($m_{L}=m_{R}=m_{C}=1$) and (b) the mass-graded system
($m_{L}=1$, $m_{R}=10$, $m_{C}=5.5$). In both panels, 
$N_{L}=N_{R}=10$, $N_{C}=64$, total system size $N_{\rm tot}=84$, $T_{+}=9.5$,
$T_{-}=0.5$, $\gamma_R=1$.}
\label{fig:onsite}
\end{figure}

In Fig.~\ref{fig:onsite}(a) we show that, even in the absence of mass
gradient ($m_L=m_R=m_C$) a moderately high rectification, 
$f_r\approx 200$$\%$, is possible when the needed asymmetry in 
the system is provided by the on-site potential, that is,
$\gamma_L/\gamma_R\gg 1$.
If we add large asymmetry for the on-site potential,
on top of a
mass-graded system, the rectification can become very large. 
For instance, in Fig.~\ref{fig:onsite}(b) we obtain 
$f_r\approx 3700$$\%$ for $\gamma_L=10 \gamma_R$ and $m_R=10 m_L$.

%%%%%%%%%%%%%%%%%%%%%%%%%%%%%%%%%%%%%%%%%%%%%%%%%%% figure 6

%%%%%%%%%%%%%%%%%%%%%%%%%%%%%%%%%%%%%%%%%%%%%%% 

\begin{figure}[!]
\includegraphics[viewport=0.0 -50 950 437.2,scale=0.3]{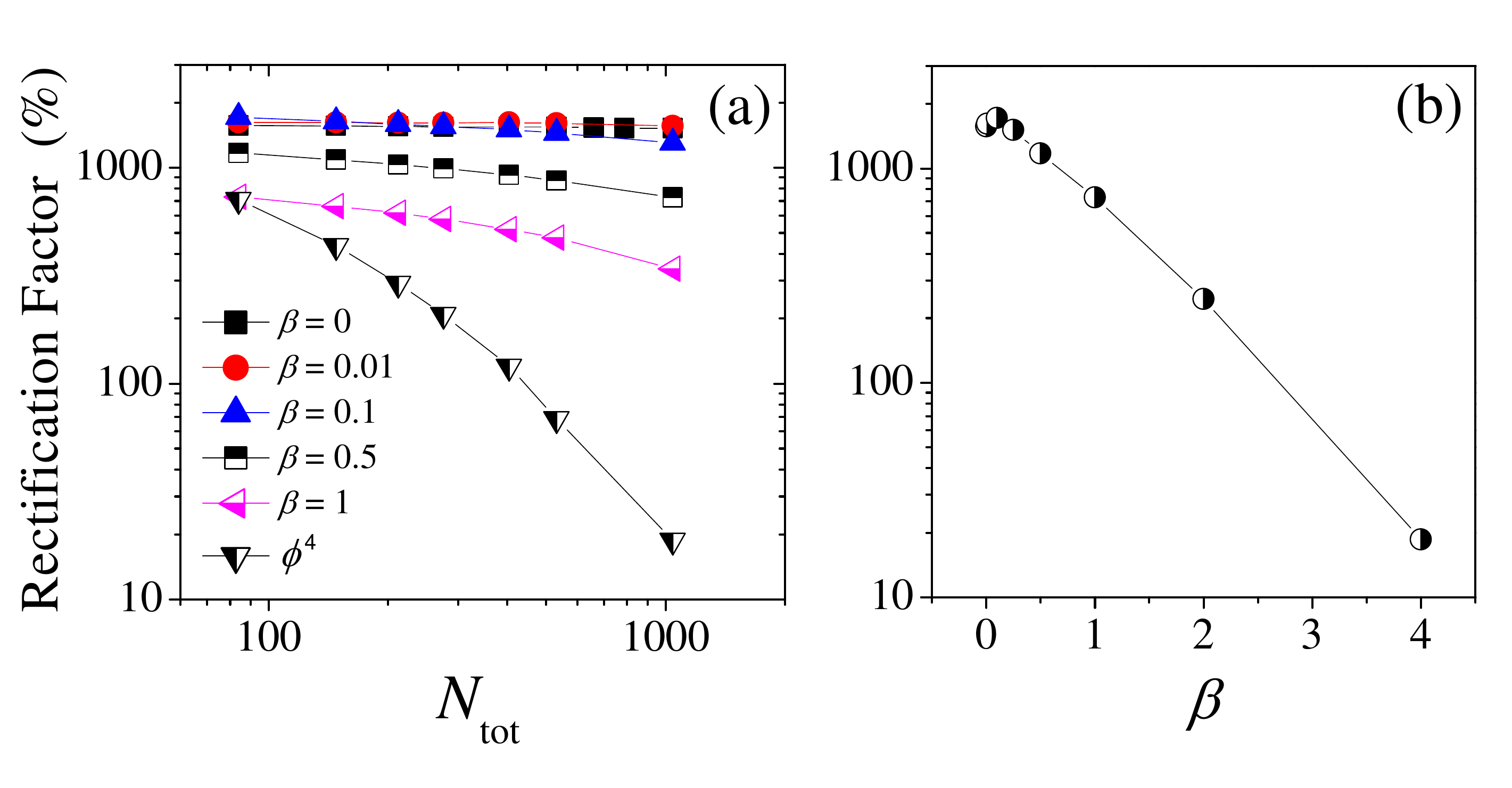} \vskip-.2cm  
\vspace*{-0.4cm} \caption{(Color online)
Effect of the anharmonicity (FPU-$\beta$ model) of the central channel 
on the rectification. (a) Rectification factor versus system size
for different values of $\beta$ ($\beta=0$ corresponds to
the ballistic, harmonic channel); the case of a diffusive, $\phi^4$ 
central channel is also shown for a comparison. 
(b) Rectification factor versus anharmonicity strength for 
size of the central channel $N_{C}=64$.
In both panels, $N_{L}=N_{R}=10$, $m_{L}=1$,
$m_{R}=10$, $m_{C}=4.5$, $T_{+}=9.5$, $T_{-}=0.5$, 
$\gamma_{L}=\gamma_{R}=1$.}
\label{fig:FPU}
\end{figure}

We finally investigate the robustness of the rectification when 
the central channel is not ballistic as in the ideal case, but 
anharmonicity is present, modeled by the Fermi-Pasta-Ulam (FPU) 
Hamiltonian FPU-$\beta$. That is, we add to the Hamiltonian for 
the $N_C$ particles in the spacer the anharmonic term 
$\sum_i \frac{\beta}{4}(q_{i}-q_{i-1})^{4}$. As a consequence
the channel is sub-ballistic \cite{Leprirev,Dharrev} and, as shown in Fig.~\ref{fig:FPU}(a),
the rectification factor decays with the system size. 
However, the decay is much slower than for a diffusive, $\phi^4$
central channel, also shown for a comparison in the same figure. 
We also show in Fig.~\ref{fig:FPU}(b) that, for a given system 
size, rectification remains large up to some value of the 
nonlinearity parameter $\beta$. For instance, we can see that 
for $N_{\rm tot}=84$ we have $f_r\gtrsim 1000\%$ for $\beta\lesssim 0.5$.

%%%%%%%%%%%%%%%%%%%%%%%%%%%%%%%%%%%%%%%%%%%%%%%%%%% figure 7

\textit{Conclusions.}
To summarize, we have discussed a model of a thermal diode,
consisting of two nonlinear systems with different masses and/or 
different on-site potentials, coupled by a ballistic thermal conductor.
The ballistic channel removes the size dependence of the rectification factor,
thus opening new possibilities of achieving high rectification factors in 
experimental devices. 

We have also checked that the same main features of our system
(high and size-independent rectification) can be obtained in 
another model, with a ballistic channel (harmonic lattice) connected
to a single mass-graded nonlinear ($\phi^4$) lead, consisting of
$N_R$ particles with masses increasing from $m_{\rm min}$ to 
$m_{\rm max}$. This shows the conceptual simplicity and 
flexibility of models based on 
mass gradient plus a ballistic channel for thermal transport. 

We expect that our results might be exploited to design a realistic thermal 
diode. In a recent work \cite{wang17}, it was demonstrated experimentally thermal rectification in various asymmetric monolayer graphene nanostructures. 
A thermal rectification factor as large as 26\% is reported in a defect-engineered asymmetric monolayer graphene with nanopores on one side. 
However, as demonstrated by molecular dynamics simulations in the same work, the thermal rectification factor would decay rapidly with system size. 
It is thus interesting if one could have possible experimental implementations considering the ballistic spacer model we discussed here, to prevent the fast decay of rectification with system size. Since graphene has very high thermal conductivity \cite{g1,g2,g3,g4}, one could consider graphene as quasi-ballistic spacer, and apply asymmetry on its two ends, where one could consider 
several asymmetric ingredients, such as defects, geometry, mass-loading, chemical functionalization, substrate couplings, mechanical strains, etc. One could also consider connecting graphene to heavier leads, such as Molybdenum disulfide (MoS$_{\rm 2}$) \cite{MoS2} or Tungsten diselenide (WSe$_{\rm 2}$) \cite{WSe2}, creating van der Waals heterostructures \cite{vdw1,vdw2} with tunable interlayer coupling strength (anharmonicity) through stress or mechanical strain. Besides graphene, one could also consider other high thermal conductivity materials, such as carbon nanotube \cite{cnt1,cnt2,cnt3,cnt4} or carbyne \cite{cb1,cb2}, as quasi-ballistic spacer, and create asymmetry on top of them to build efficient thermal diodes.

\textit{Acknowledgments:} We acknowledge the support by the CINECA project
Nanostructures for Heat  Management and Thermoelectric Energy Conversion.


\begin{thebibliography}{10}
\bibitem{Casati1} M. Terraneo, M. Peyrard M., and G. Casati, Phys. Rev.
Lett., \textbf{88}, 094302 (2002).
\bibitem{Casati2} 
B. Li, L. Wang, and G. Casati,
Phys. Rev. Lett. \textbf{93}, 184301 (2004).
\bibitem{Casati3} 
B. Li, L. Wang, and G. Casati,
Appl. Phys. Lett. \textbf{88}, 143501 (2006).
\bibitem{BaowenRMP}
N. Li, J. Ren, L. Wang, G. Zhang, P. H\"{a}nggi, and B. Li,
Rev. Mod. Phys. \textbf{84}, 1045 (2012).
\bibitem{Benenti2016} G. Benenti, G. Casati, C. Mej\'{\i}a-Monasterio, and M. Peyrard, \textit{From thermal rectifiers to thermoelectric devices}, in \textit{Thermal transport in low dimensions}, S. Lepri (Ed.), Lecture Notes in Physics \textbf{921} (Springer, 2016).
\bibitem{majumdar}
C. W. Chang, D. Okawa, A. Majumdar, and A. Zettl, 
Science \textbf{314}, 1121 (2006).
\bibitem{terasaki1}
W. Kobayashi, Y. Teraoka, and I. Terasaki, 
Appl. Phys. Lett. \textbf{95}, 171905 (2009).
\bibitem{terasaki2}
D. Sawaki, W. Kobayashi, Y. Morimoto, and I. Terasaki,  
Appl. Phys. Lett. \textbf{98}, 081915 (2011).
\bibitem{terasaki3}
W. Kobayashi, D. Sawaki, T. Omura, T. Katsufuji, Y. Moritomo, 
and I. Terasaki, Appl. Phys. Express \textbf{5}, 027302 (2012). 

\bibitem{NJP11}
M. Schmotz, J. Maier, E. Scheer, and P. Leiderer, New J. Phys. 
\textbf{13}, 113027 (2011).

\bibitem{tian12}
H. Tian, D. Xie, Y. Yang, T. L. Ren, G. Zhang, Y. F. Wang, 
C. J. Zhou, P. G. Peng, L. G. Wang, and L. T. Liu,
Sci. Rep. \textbf{2}, 523 (2012).
\bibitem{wang17}
H. Wang, S. Hu, K. Takahashi, X. Zhang, H. Takamatsu, and J. Chen,
Nature Comm. \textbf{8}, 15843 (2017).
\bibitem{giazotto}
We do not discuss here the thermal rectification of photonic or electronic
currents; in particular, large rectification of electronic heat current
in a hybrid device combining normal metals tunnel-coupled to superconductors,
was reported in M. J. Martinez-P\'erez, A. Fornieri, and F. Giazotto,
Nat. Nanotechnol. \textbf{10}, 303 (2015).
\bibitem{pereira2013}
E. Pereira and R. R. \'Avila, 
Phys. Rev. E \textbf{88}, 032139 (2013).
\bibitem{shunda2015}
S. Chen, E. Pereira, and G. Casati,
EPL \textbf{111}, 30004 (2015).
\bibitem{Dharrev} A. Dhar, Adv. Phys. \textbf{57}, 457 (2008).

\bibitem{destri}
D. Boyanovsky, C. Destri, and H. J. de Vega, 
Phys. Rev. D \textbf{69}, 045003 (2004).
\bibitem{baowen}
N. Li and B. Li, Phys. Rev. E \textbf{87}, 042125 (2013).

\bibitem{Leprirev}
S. Lepri, R. Livi, and A. Politi,
Phys. Rep. \textbf{377}, 1 (2003).

\bibitem{g1}
A. A. Balandin, S. Ghosh, W. Bao, I. Calizo, D. Teweldebrhan, F. Miao, and C. N. Lau, Nano Lett. \textbf{8}, 902 (2008).

\bibitem{g2}
A. A. Balandin, Nat. Mater. \textbf{10}, 569 (2011).

\bibitem{g3}
L. F. C. Pereira and D. Donadio, Phys. Rev. B \textbf{87}, 125424 (2013).

\bibitem{g4}
X. Xu, L. F. C. Pereira, Y. Wang, J. Wu, K. Zhang, X. Zhao, S. Bae, C. T. Bui, R. Xie, J. T. L. Thong, B. H. Hong, K. P. Loh, D. Donadio, B. Li, and B. \"{O}zyilmaz, Nature Comm. \textbf{5}, 3689 (2014).

\bibitem{MoS2}
K. F. Mak, C. Lee, J. Hone, J. Shan, and T. F. Heinz, Phys. Rev. Lett. \textbf{105}, 136805 (2010).

\bibitem{WSe2}
C. Chiritescu, D. G. Cahill, N. Nguyen, D. Johnson, A. Bodapati, P. Keblinski, and P. Zschack, Science \textbf{315}, 351 (2007).

\bibitem{vdw1}
A. K. Geim and I. V. Grigorieva, Nature \textbf{499}, 419 (2013).

\bibitem{vdw2}
K. S. Novoselov, A. Mishchenko, A. Carvalho, and A. H. C. Neto, Science \textbf{353}, 461 (2016).

\bibitem{cnt1}
P. Kim, L. Shi, A. Majumdar, and P. L. McEuen, Phys. Rev. Lett. \textbf{87}, 215502 (2001).

\bibitem{cnt2}
E. Pop, D. Mann, Q. Wang, K. Goodson, and H. Dai, Nano Lett. \textbf{6}, 96 (2006).

\bibitem{cnt3}
D. Donadio and G. Galli, Phys. Rev. Lett. \textbf{99}, 255502 (2007).

\bibitem{cnt4}
L. F. C. Pereira and D. Donadio, Phys. Rev. B \textbf{87}, 125424 (2013).

\bibitem{cb1}
M. Wang and S. Lin, Sci. Rep. \textbf{5}, 18122 (2015).

\bibitem{cb2}
L. Shi, P. Rohringer, K. Suenaga, Y. Niimi, J. Kotakoski, J. C. Meyer, H. Peterlik, M. Wanko, S. Cahangirov, A. Rubio, Z. J. Lapin, L. Novotny, P. Ayala, and T. Pichler, Nat. Mater. \textbf{15}, 634 (2016).





\end{thebibliography}
\end{document}